\documentclass[preprint,showpacs,preprintnumbers,amsmath,amssymb]{revtex4}
\usepackage{bm}
\usepackage{amsmath}
\usepackage{amsfonts}
\usepackage{graphicx}
\usepackage{color}

\newcommand{\nn}{\nonumber}

\newcommand{\zd}{\delta}

\newcommand{\zs}{\sigma}

\newcommand{\ze}{\varepsilon}
\newcommand{\zb}{\beta}
\newcommand{\zg}{\gamma}
\newcommand{\zl}{\lambda}

\newcommand{\zr}{\rho}

\newcommand{\zh}{\eta}

\newcommand{\dif}{\; \textrm d }

\newcommand{\dpp}[2]{\frac{  \partial #1 }{  \partial #2   } }
\newcommand{\dpt}[2]{\frac{  \dif #1 }{  \dif #2   } }

\begin{document}

\title{Ultrafast Magnetization Dynamics in Diluted
Magnetic Semiconductors}

\author{O. Morandi$^{1}$, P. -A. Hervieux$^{2}$, G. Manfredi$^{2}$}
\affiliation{$^1$INRIA Nancy Grand-Est and Institut de Recherche
en Math{\'e}matiques Avanc{\'e}es, 7 rue Ren{\'e} Descartes, F-67084
Strasbourg, France
 \\ $^2$Institut de Physique et Chimie des Mat{\'e}riaux de Strasbourg, 23 rue du Loess, F-67037
Strasbourg, France}

\date{\today}

\begin{abstract}
We present a dynamical model that successfully explains the
observed time evolution of the magnetization in diluted magnetic
semiconductor quantum wells after weak laser excitation. Based on
the pseudo-fermion formalism and a second order many-particle
expansion of the exact $p-d$ exchange interaction, our approach
goes beyond the usual mean-field approximation. It includes both
the sub-picosecond demagnetization dynamics and the slower
relaxation processes which restore the initial ferromagnetic order
in a nanosecond time scale. In agreement with experimental
results, our numerical simulations show that, depending on the
value of the initial lattice temperature, a subsequent enhancement
of the total magnetization may be observed within a time scale of
few hundreds of picoseconds.
\end{abstract}


\maketitle

\section{Introduction}
Ultrafast light-induced magnetization dynamics in ferromagnetic
films and in Diluted Magnetic Semiconductors (DMS) is today a very
active area of research. From the observation of the ultrafast
dynamics of the spin magnetization in nickel films
\cite{Beaurepaire_96} and the analogous processes in ferromagnetic
semiconductors \cite{Ohno_98}, special interest has been devoted
to the development of dynamical models able to mimic the time
evolution of the magnetization on both short and long time scales.
In III-V ferromagnetic semiconductors such as GaMnAs and InMnAs a
small concentration of Mn ions is randomly substituted to cation
sites so that the Mn-Mn spin coupling is mediated by the hole-ion
$p-d$ exchange interaction, allowing the generation of a
ferromagnetic state with a Curie temperature of the order of 50 K
\cite{Dietl_00}. The magnetism can therefore be efficiently
modified by controlling the hole density through doping or by
excitation of electron-hole pairs with a laser pulse. In
particular, unlike metals, in a regime of strong laser excitation
total demagnetization can be achieved \cite{Wang_05}.

In the Zener model \cite{Zener_51}, which was originally developed
to describe the magnetism of transition metals, the $d$ shells of
the Mn ions are treated as an ensemble of randomly distributed
impurities with spin $5/2$ surrounded by a hole gas or an electron
gas. Unlike ferromagnetic metals, III-Mn-V ferromagnetic
semiconductors offer the advantage of providing a clear
distinction between localized Mn impurities and itinerant
valence-band hole spins, thus allowing the basic assumptions of
the Zener theory to be satisfied. Based on this hypothesis, a few
mean-field models have been successfully applied for modelling the
ground-state properties of DMS nanostructures. In particular,
within the framework of the spin-density-functional theory at
finite temperature, relevant predictions of the Curie temperature
have been obtained \cite{Lee_00,Kim_06}. Ultrafast demagnetization
in DMS is a phenomenon where the $p-d$ exchange interaction causes
a flow of spin polarization and energy from the Mn impurities to
the holes, which is subsequently converted to orbital momentum and
thermalized through spin-orbit and hole-hole interactions
\cite{Tserkovnyak_04}. Since energy and spin polarization transfer
is a many-particle effect, the mean-field Zener approach cannot
provide a satisfying explanation of the ultrafast demagnetization
regime that has been observed in DMS \cite{Wang_05,Wang_07}.

A phenomenological approach able to take into account this energy
flux was given in \cite{Beaurepaire_96, Konig_00} where a model
based on three temperatures was derived. More recently, a study of
the coupling of the electromagnetic laser field with the hole gas
revealed the possibility of an ultrafast demagnetization during
the femtosecond optical excitation, due to light-hole entanglement
\cite{Chovan_06}. A model capable of describing the dynamics of
carrier-ion spin interactions is provided in \cite{Cywiski_07,
Wang_08}. This model generalizes the stationary theory of
\cite{Konig_00} and takes into account the picosecond
demagnetization evolution which occurs in a strong excitation
regime, but neglects the slow-in-time evolution of the spin
dynamics. The mean $p-d$ interaction is averaged out over the
randomly distributed positions of the Mn ions.


In this work we derive a dynamical model based on a many-particle
expansion of the $p-d$ exchange interaction based on the
pseudo-fermion method. This formalism, originally developed by
Abrikosov to deal with the Kondo problem \cite{Abrikosov_65},
introduces unphysical states in the Hilbert space for which
impurity sites are allowed to be multiply occupied. Following the
work of Coleman \cite{Coleman_83}, a suitable limit procedure is
applied to our dynamical model in order to recover the correct
physical description of the magnetic impurities.

Our approach extends the Zener model beyond the usual mean-field
approximation. It includes both the subpicosecond demagnetization
dynamics and the slower cooling processes that restore the initial
ferromagnetic order (which is achieved in a ns time scale).
Moreover, in agreement with recent experimental results
\cite{Wang_07} our simulations show that, depending on the initial
lattice temperature, a subsequent enhancement of the total
magnetization is observed within a time scale of 100 ps.

\section{Pseudo-fermion formalism}
We consider a volume $V$ containing $N^h V$ holes with spin
$S^h=1/2$ strongly coupled by spin-spin interaction with $N^M V$
randomly distributed Mn impurities with spin $S^M=5/2$. We assume
that the exchange interaction between localized ions and heavy
holes dominates over both the short-range antiferromagnetic $d-d$
exchange interaction between the ions and the $s-d$ exchange
interaction between electrons in the conduction band and Mn ions
(typical values for the $s-d$ and the $p-d$ interactions in a GaAs
are 0.1 eV and 1 eV respectively \cite{Winkler_book}).
Furthermore, electron-hole radiative recombination, carrier-phonon
interactions, and interactions leading to the hole spin-relaxation
in the hole gas are included phenomenologically. The time
evolution of the system is governed by the Hamiltonian
\begin{equation*}
\mathcal{H}=\sum_{k,s}   \ze_{k,s} {a}_{k,s}^\dag {a}_{k,s}
+\mathcal{H}_{pd} \;,
\end{equation*}
where $a^\dag_{k,s}$ ($a_{k,s}$) is the creation (annihilation) operator of a hole with spin
projection $s$ and quasi-momentum $k$.
In the parabolic band approximation the kinetic energy of the
holes reads $\ze_{k,s}= E^h - \frac{\hbar^2 k^2}{2 m^*} $ where
$E^h$ is the valence band edge.
The Kondo-like exchange interaction $\mathcal{H}_{pd}$ is given by
\begin{equation}\label{H_pd sq_1}
\mathcal{H}_{pd}= \frac{\zg}{V} \sum \mathbf{J}_{m',m}\cdot
\boldsymbol{\zs}_{s',s} \left( {b}_{\zh,m'}^\dag b_{\zh,m}
{a}_{k',s'}^\dag {a}_{k,s}\right) e^{i
(k'-k)R_{\zh}} \nn
\end{equation}
where the sum is extended over all indices, $\zg$ is the $p-d$
coupling constant, and $\boldsymbol{\zs}$, $\mathbf{J}$ are the
spin matrices related to $S^h$ and $S^M$ respectively. The ion
spin operator is represented in the pseudo-fermion formalism
\cite{Abrikosov_65,Coleman_83} in which $b^\dag_{\zh,m}$
($b_{\zh,m}$) denotes the creation (annihilation) operator of a
pseudo-fermion with spin projection $m$ and spatial position
$R_\zh$.

The $\mathcal{H}_{pd}$ Hamiltonian reproduces the correct ion-hole
exchange interaction provided that the ion sites are singly
occupied, i.e., $\hat{n}_\zh=\sum_{m=-S^M}^{S^M} b^\dag_{\zh,m}
b_{\zh,m} = 1$ $\forall \zh$. Following
\cite{Abrikosov_65,Coleman_83} this constraint may be taken into
account by adding a "fictitious" ionic chemical potential
\begin{eqnarray*}
\mathcal{H}^\zl &=& \sum_{\zh } \zl_\zh\;\hat{n}_\zh
\end{eqnarray*}
to the original Hamiltonian and letting $\zl_\zh$ go to infinity
at the end of the calculation.

The grand-canonical expectation value of a pseudo-fermion operator
$\mathcal{A}$ related to the total Hamiltonian $\mathcal{H}+
\mathcal{H}^\zl$ reads
\begin{eqnarray*}
\left\langle \mathcal{A} \right\rangle_{\zl}&=&\frac{1}{\mathcal{Z}_{\zl}}\textrm{Tr}
\left\{\zr_\mathcal{H} e^{-\zb  \sum_\zh \zl_\zh \hat{n}_\zh } \mathcal{A}  \right\}\\
&=&\frac{1}{\mathcal{Z}_{\zl}}  \sum_{\left\{  {n}_\zh^{m}
\right\}_r }\left\langle  \; n_\zh^{m} \; \right|\zr_\mathcal{H}
e^{-\zb  \sum_\zh \zl_\zh \hat{n}_\zh } \mathcal{A}    \left|  \;
n_\zh^{m} \; \right\rangle\; ,
\end{eqnarray*}
where $\mathcal{Z}_\zl  =  \textrm{Tr}\left\{\zr_\mathcal{H}
e^{-\zb  \sum_\zh \zl_\zh \hat{n}_\zh } \right\}$,
$\zr_\mathcal{H}= e^{-\zb \mathcal{H}}$, and $\zb=1/k_B T^h$, with
$k_B$ the Boltzmann constant and $T^h$ the hole temperature.
$\left\{  {n}_{{\zh}}^{{m}} \right\}_r= \left\{
{n}_1^{1},\ldots,{n}_1^{(2S^M+1)},\ldots,{n}_r^{(2S^M+1)} \right\}
$ denotes all possible occupation numbers $n_\zh^{k}$ ($=0$ or
$1$) for $r$ ion sites. Since each site has $(2S^M+1)$ available
pseudo-fermion states, the system will contain at most $(2S^M+1) r
$ pseudo-particles. The correct expectation value of the operator
$\mathcal{A}$ is obtained using the limit $\zl_\zh \rightarrow
\infty$ \cite{Coleman_83}
\begin{eqnarray}
\left\langle \mathcal{A}  \right\rangle_{\infty}
&=&\frac{1}{\mathcal{Z}_\infty}\lim_{\left\{ z_\zh\right\}
\rightarrow 0} \frac{\partial^r  \left[ \left\langle \mathcal{A}
\right\rangle_{\zl} \mathcal{Z}_\zl \right]}{\partial
z_1\cdots\partial z_r} \;, \label{pse lim}
\end{eqnarray}
where $\mathcal{Z}_\infty = \lim_{\left\{ z_\zh\right\}
\rightarrow 0} \frac{\partial^r \mathcal{Z}_\zl}{\partial
z_1\cdots\partial z_r}$ and $z_\zh=e^{-\zb\zl_\zh}$.

In the next section, we will show that the time-evolution of the
spin of the ion-hole system may be expressed in terms of the
expectation value of the pseudo-fermion operator $b_{\zh,m}^\dag
b_{\zh,m} (1- b_{\zh,m'}^\dag b_{\zh,m'})$ with $m\neq m'$ and
evaluated in the mean magnetic field $\mathbf{S}$ generated by the
holes. We have the general relationship (which also applies when
the system is driven far from equilibrium)
\begin{eqnarray}
\lim_{\left\{ \zl_\zh \right\} \rightarrow \infty} \left\langle
b_{\zh,m}^\dag b_{\zh,m} (1- b_{\zh,m'}^\dag b_{\zh,m'})
\right\rangle_{\zl} =\lim_{\left\{ \zl_\zh \right\} \rightarrow
\infty} \left\langle b_{\zh,m}^\dag b_{\zh,m}
\right\rangle_{\zl}\; . \label{pse_lim_bb}
\end{eqnarray}
When the system approaches thermal equilibrium, the quantity
$\left\langle b_{\zh,m}^\dag b_{\zh,m} \right\rangle_{\infty}$
becomes the usual spin thermal distribution. Using Eq. (\ref{pse
lim}) we obtain
\begin{eqnarray}\label{bb eq}
\left\langle b_{\zh,m}^\dag b_{\zh,m}  \right\rangle_{\infty} &=&
\widetilde{\mathcal{Q}}\; \frac{e^{\zb m \gamma \mathbf{S}}
}{\mathcal{Z}_\infty }\;,
\end{eqnarray}
where $\mathcal{Z}_\infty = \widetilde{\mathcal{Q}}  \;
\frac{\sinh\left[\frac{\zb  \gamma \mathbf{S}  }{2}\left(2 S^M +1
\right) \right] }{\sinh \left(\frac{\zb  \gamma \mathbf{S} }{2}
\right) }$ and
$\widetilde{\mathcal{Q}}=\left.\mathcal{Q}\right|_{{n}_{\zh'}^{m}=0,1;
\sum_m {n}_{\zh'}^{m}=1}$ with $\mathcal{Q}=\prod_{m,\zh'\neq\zh}
e^{-\zb m \gamma \mathbf{S} \; n_{\zh'}^m}$.

In order to derive Eq. (\ref{bb eq}), we have used
\begin{eqnarray*}
 \lim_{\left\{ z_\zh \right\} \rightarrow 0} \frac{\partial^r
}{\partial z_1\cdots\partial z_r}\textrm{Tr}\left\{\zr_\mathcal{H}
e^{-\zb  \sum_{\zh'}  \zl_{\zh'} \hat{n}_{\zh'}    }
\hat{n}_{\zh}^m \right\} = \nn \widetilde{\mathcal{Q}}
\sum_{{n}_\zh^{m}=0,1; \sum_m {n}_\zh^{m}=1}  n_\zh^m e^{-\zb
\gamma \mathbf{S}  \;m n_\zh^m}=\widetilde{\mathcal{Q}}\; e^{-\zb
\gamma \mathbf{S}\;m }\label{lim_z_zero}
\end{eqnarray*}

with $\zr_\mathcal{H}  = e^{-\zb  \gamma \mathbf{S} \sum_{\zh,m} m
\hat{n}_\zh^m}$ and $\hat{n}^m_\zh= b_{\zh,m}^\dag b_{\zh,m}$.

\section{Time evolution model}
The Heisenberg equations of motion lead to a hierarchy of time
evolution equations for the mean densities $n^h_s =
\frac{1}{N^h}\sum_k \langle a_{k,s}^\dag a_{k,s}\rangle_\infty$
and $n^M_m = \frac{1}{N^M} \sum_\zh \langle b_{\zh,m}^\dag
b_{\zh,m}\rangle_\infty$
\begin{eqnarray}\label{evol fin a dag_a}
\dpt{[\sum_k \langle a_{k,s}^\dag a_{k,s}\rangle_\lambda ]}{t}&=&
N^h N^M \sum_{m_1} \mathcal{W}_{s,s,m_1,m_1}\\
\dpt{[\sum_\zh \langle b_{\zh,m}^\dag
b_{\zh,m}\rangle_\lambda]}{t}&=&\label{evol fin b_dag_b} N^M N^h
\sum_{s_1} \mathcal{W}_{s_1,s_1,m,m}\;,
\end{eqnarray}
with
\begin{eqnarray}
\mathcal{W}_{s,s,m,m}= \sum_{s'_1,m'_1}\left( \mathbf{J}_{m'_1,m}
\cdot \boldsymbol{\zs}_{s'_1,s} \;
\widetilde{\mathcal{C}}_{m'_1,m,s_1',s} - \mathbf{J}_{m,m'_1}
\cdot \boldsymbol{\zs}_{s,s_1'} \;
\widetilde{\mathcal{C}}_{m,m_1',s,s_1'}\right) \;. \label{Wscat}
\end{eqnarray}
In the last equation, the mean correlation function reads
\begin{equation}
\widetilde{\mathcal{C}}_{m',m_1,s_1',s_1} = - \frac{i}{ \hbar
}\frac{ \zg}{ V N^h N^M}\sum_{\zh,k_1,k_1'}
\mathcal{C}_{m',m_1,s_1',s_1}^{\zh,\zh,k_1',k_1}e^{i
(k_1-k_1')R_\zh} \;,
\end{equation}
where ${\mathcal{C}_{m',m,s',s}^{\zh',\zh,k',k}}  = \langle
{b}_{\zh',m'}^\dag b_{\zh,m} {a}_{k',s'}^\dag a_{k,s}\rangle_\zl$.
The time evolution equation of this quantity is given by
\begin{eqnarray}\label{evol C}
i\hbar\dpt{{\mathcal{C}}_{\mathbf{m}',\mathbf{m},\mathbf{s}',\mathbf{s}}}{t}&=&
\Delta E_{MF}\;
{\mathcal{C}}_{\mathbf{m}',\mathbf{m},\mathbf{s}',\mathbf{s}}
\\ \nonumber
&+& \frac{\zg}{V}
\sum_{\mathbf{m}'_1,\mathbf{m}_1,\mathbf{s}_1,\mathbf{s}'_1}
\zd_{\zh,\zh'} \mathbf{J}_{m'_1,m_1}\cdot
\boldsymbol{\zs}_{s_1',s_1} \left\langle\mathcal{B}
\mathcal{A}-\mathcal{A}^t \mathcal{B}^t \right\rangle_\zl e^{i
(k_1'-k_1)R_{\zh_1}}
\end{eqnarray}
where the compact notations $\mathbf{m}\equiv (\zh,m)$,
$\mathbf{s} \equiv (k,s)$, $\mathcal{B} \equiv
{b}_{{\mathbf{m}}_1'}^\dag b_{{\mathbf{m}}_1}
b^\dag_{{\mathbf{m}}'} b_{\mathbf{m}}$, $\mathcal{B}^t \equiv
b^\dag_{{\mathbf{m}}'} b_{\mathbf{m}} {b}_{{\mathbf{m}}_1'}^\dag
b_{{\mathbf{m}}_1}$, $\mathcal{A} \equiv
{a}_{{\mathbf{s}}_1'}^\dag{a}_{{\mathbf{s}}_1}a^\dag_{{\mathbf{s}}'}a_{\mathbf{s}}$,
$\mathcal{A}^t \equiv a^\dag_{{\mathbf{s}}'}a_{\mathbf{s}}
{a}_{{\mathbf{s}}_1'}^\dag {a}_{{\mathbf{s}}_1}$ have been
employed.

The mean-field contribution to the total energy is given by $
\Delta E_{MF}=  \zg \left[\left(s'-s\right)   \mathbf{M} +
\left(m'-m\right) \mathbf{S}\right] $ where $\mathbf{M}= N^M
\sum_{m=-S^M}^{S^M} m \; n^M_m $ and $\mathbf{S}= N^h
\sum_{s=-S^h}^{S^h} s \; n^h_s $ are the mean magnetic field
generated by the ions and by the holes respectively.

The use of Eq. (\ref{evol C}) combined with Eqs. (\ref{evol fin a
dag_a})-(\ref{evol fin b_dag_b}) leads to a non-Markovian time
evolution of the macroscopic dynamical variables such as the
density and the magnetization. By assuming an instantaneous
spin-spin interaction, the Markov approximation can be easily
recovered. For further details about the justification of the
Markovian approximation in a DMS excited by a laser pulse we refer
to \cite{Cywiski_07}.

By using the Dirac identity \cite{Haug_Koch_book}
$\int_{-\infty}^t
 e^{-i\ze(t-t')/\hbar} \dif t' =-\pi \hbar \zd\left(\ze  \right)- i\hbar \mathcal{P}
 \frac{1}{\ze}$ where $\mathcal{P}$ denotes the principal value, the integration
of Eq. (\ref{evol C}) with respect to the time leads to
\begin{eqnarray}
 {\mathcal{C}}_{\mathbf{m}',\mathbf{m},\mathbf{s}',\mathbf{s}} \nn&=&
-i \pi \frac{\zg}{V}
\sum_{\mathbf{m}_1,\mathbf{m}_1',\mathbf{s}_1,\mathbf{s}_1' } \zd \left(\Delta \mathbb{E}_{MF} \right)
\\&\times&
 \mathbf{J}_{m'_1,m_1}\cdot \boldsymbol{\zs}_{s'_1,s_1} \left\langle \mathcal{B}\mathcal{A} -\mathcal{B}^t\mathcal{A}^t
\right\rangle_\zl e^{i \left[(k_1'-k_1)R_{\zh_1}+
(k'-k)R_\zh\right]}\label{C mal mark} \; ,
\end{eqnarray}
where $\Delta \mathbb{E}_{MF}= \ze_{k'}- \ze_{k} + \Delta E_{MF}$.

Since the matrix operators $\mathbf{J} \cdot \boldsymbol{\zs}$ are
real, it is clear from Eq. (\ref{Wscat}) that the imaginary part
gives no contribution to the equation of motion.

The many-particle expansion of the correlation function
$\widetilde{\mathcal{C}}$ allows us to express Eq. (\ref{C mal
mark}) in terms of the single-particle density matrix elements
$n_{s}^h$ and $n_{m}^M$. By using the commutation rules of the
creation and annihilation operators we obtain
\begin{eqnarray*}
\left\langle \mathcal{B}\mathcal{A} -\mathcal{B}^t\mathcal{A}^t
\right\rangle_\zl &=& \zd_{{\mathbf{m}}_1,{\mathbf{m}}'} \zd_{{\mathbf{m}},{\mathbf{m}}_1'} \zd_{{\mathbf{s}}_1,{\mathbf{s}}'} \zd_{{\mathbf{s}}_1',{\mathbf{s}}}\\&& \left\langle
 \left(b^\dag_{{\mathbf{m}} }b_{\mathbf{m}}\;-
b^\dag_{{\mathbf{m}}'}b_{{\mathbf{m}}'}\; \right)   a^\dag_{\mathbf{s}}a_{\mathbf{s}}\;  \left(  1 -
a^\dag_{{\mathbf{s}}'} a_{{\mathbf{s}}'} \right) \right.\\
&+& \left. b^\dag_{\mathbf{m}}b_{\mathbf{m}}\;  \left(  \ 1 -
b^\dag_{{\mathbf{m}}'} b_{{\mathbf{m}}'} \right)  \left(
a^\dag_{{\mathbf{s}} }a_{\mathbf{s}}\;-
a^\dag_{{\mathbf{s}}'}a_{{\mathbf{s}}'}\; \right)
\right\rangle_\zl \;.
\end{eqnarray*}
Furthermore, as a closure hypothesis, we have assumed that the
non-diagonal matrix elements of the density-like operators
$a^\dag_{{\mathbf{s}'} }a_{\mathbf{s}}$ and $b^\dag_{{\mathbf{m}'}
}b_{\mathbf{m}}$ with respect to the indexes $\zh$ and $k$ vanish.
From the above approximations and using the definition (\ref{evol
fin a dag_a}) and Eq. (\ref{C mal mark}) we get
\begin{eqnarray}
\sum_{m_1} \mathcal{W}_{s,s,m_1,m_1} &=& \frac{2 \pi}{\hbar N^S
N^M}\left(\frac{\zg}{V}\right)^2\sum_{s_1,m'_1,m_1}
\mathbf{J}_{m'_1,m_1} \cdot \boldsymbol{\zs}_{s ,s_1}
\mathbf{J}_{m_1,m_1'} \cdot \boldsymbol{\zs}_{s_1,s} \nn \\&&
\sum_{ k,k',\zh }   \zd\left(\Delta \mathbb{E}_{MF} \right)
\left(\Pi^\zl_{m_1,m_1',s,s_1} -\Pi^\zl_{m_1',m_1,s_1,s} \right)
\label{W fin}
\end{eqnarray}
where,
\begin{eqnarray}
\Pi^\zl_{m_1,m_1',s,s_1} &=& \sum_{ k,k_1,\zh } \left\langle
\left(  1 - b^\dag_{\zh,{{m}}_1} b_{\zh,{{m}}_1} \right)
b^\dag_{\zh,{{m}}_1'} b_{\zh,{{m}}_1'}\right.  \left.
a^\dag_{k,{s}}a_{k,{s}}\;  \left(  1 - a^\dag_{k_1,{{s}}_1}
a_{k_1,{{s}}_1} \right)\right\rangle_\zl \;.
\end{eqnarray}
A similar expression can be found for $\sum_{s_1}
\mathcal{W}_{s_1,s_1,m,m}$ in Eq. (\ref{evol fin b_dag_b}). By
using Eq. (\ref{pse_lim_bb}) we recover the fermionic limit of
$\Pi$, namely
\begin{eqnarray}
\Pi^\infty_{m_1,m_1',s,s_1} &=& N^M n^M_{m_1'} \sum_{ k,k_1}
\left\langle a^\dag_{k,{s}}a_{k,{s}} \;  \left(  1 -
a^\dag_{k_1,{{s}}_1} a_{k_1,{{s}}_1} \right)\right\rangle_\infty
\;.\label{pse lim P}
\end{eqnarray}
In order to evaluate the time derivative of $n_s^h$, Eq. (\ref{pse
lim P}) can be solved numerically. In the following paragraphs, we
show that Eq. (\ref{pse lim P}) may actually be further
simplified. According to the Zener model the ground state of the
system can be estimated by taking into account only the mean-field
interaction between the holes and the magnetic ions. The hole gas
experiences a mean magnetic field equal to $\mathbf{M}$ and in
turn generates a mean field acting on the ions system equal to
$\mathbf{S}$. By converting the sum over $k$ and $k_1$ in Eq.
(\ref{pse lim P}) into the corresponding integral with respect to
the energy variable $E=\ze_{k}$, we obtain
\begin{eqnarray}
\frac{\Pi^\infty}{ V^2 }&=& N^M n^M_{m_1'} e^{-\frac{\Delta E_{MF}
}{k_B T^h}} \int f_a \; h \;  \zr(E)\zr(E- \Delta  E_{MF} ) \dif E
\;, \label{Xi}
\end{eqnarray}
where $f_a= \langle a^\dag_{\mathbf{s}} a_\mathbf{s}\rangle_\infty
 \left(1-\langle{a}_{\mathbf{s_1}}^\dag a_{\mathbf{s_1}}\rangle_\infty
\right)$, $h = \frac{  1+e^{\left[\zg s_1 \mathbf{M} + \ze_k
\right]/k_B T^h} }{ 1+e^{\left[\zg  s_1 \mathbf{M}
+\ze_{k_1}\right]/k_B T^h} }$ and $\zr$ denotes the hole density
of states.

In the limit $\zg\mathbf{S}\ll\zg \mathbf{M}\ll \ze_{k}$ we have
\begin{eqnarray}\label{app_P}
\frac{\Pi^\infty}{V^2} &\simeq& N^M N^h  n_{m_1'}^M \left(\frac{2
m^*}{\hbar^2}\right) \sqrt[3]{3 \pi^2 N^h} e^{-\frac{ \Delta
E_{MF}}{k_B T^h}}   n_{s}^h \left(1-{ n_{s_1}^h} \right).
\end{eqnarray}
In the next section we will validate this approximation by
comparing the time evolution of the magnetization obtained by
using either the approximate formula (\ref{app_P}) or the exact
one (\ref{pse lim P}). Finally, by inserting Eq. (\ref{app_P})
into Eq. (\ref{W fin}) we obtain
\begin{eqnarray}
\dpt{ n_{s}^h }{t} & =& 2 \xi N^M
\frac{s}{|s|}\sum_{m=-S^M}^{S^M-1}
\left(S^M-m\right)\left(S^M+m+1\right) \left(\mathcal{Z}_{m
}^{1/2,-1/2}-\mathcal{Z}_{m+1 }^{-1/2,1/2}\right)\label{fin evol nh} \\
\dpt{n^M_{m}}{t}&=& 2 \xi N^h \sum_{\zs=\pm 1} \left(S^M-\zs
m+1\right)\left(S^M+\zs m\right) \left(\mathcal{Z}_{m
}^{-\zs/2,\zs/2}-\mathcal{Z}_{m-\zs
}^{\zs/2,-\zs/2}\right)\label{fin evol nm}
\end{eqnarray}
where
\begin{eqnarray}
\mathcal{Z}_{m }^{s,s'}&=&n^M_{m}   \; n_{s}^h \left(1-{ n_{s'}^h}
\right) e^{-\frac{\Delta E_{MF}}{k_B T^h}} \;,
\end{eqnarray}
with $\xi =  2 \pi \zg^2   \frac{  m^*}{\hbar^3} \sqrt[3]{3 \pi^2
N^h}$.

\section{Spin evolution in DMS}
In order to study the time evolution of the mean magnetization of
a GaMnAs/GaAs DMS heterostructure occurring after the interaction
with a linearly polarized femtosecond laser pulse, we have applied
our time dependent model constituted of Eqs. (\ref{fin evol nh})
and (\ref{fin evol nm}). Based on the experiment of
\cite{Wang_07}, we consider a sample consisting of a 73 nm
$\textrm{Ga}_{0.925}\textrm{Mn}_{0.075}\textrm{As}$ layer
deposited on a GaAs buffer layer and a semi-insulating GaAs
substrate. The background hole density is chosen to be $10^{20}$
cm$^{-3}$. For the details of the chemical composition of the
sample we refer to \cite{Wang_07}. We assume that before the laser
is turned on, the ion-hole system is at equilibrium with the
phonon bath at the lattice temperature $T^L$, so that the ground
state can be well described by the Zener-type model described in
\cite{Kim_06}. The laser excitation generates a non-thermal
electron-hole pairs distribution. By means of the Coulomb
hole-hole interaction, the hole distribution undergoes a
quasi-instantaneous thermalization (within a few tens of
femtoseconds) towards a Fermi-Dirac distribution with temperature
$T^h$ and a spin dependent chemical potential $\mu^h_s$
\cite{Koopmans_05,Shah_book}. The increasing of the overall
temperature $T^h$  of the hole gas is determined by assuming that
the excess of energy of the hot photo-created particles (which is
estimated as a fraction of the pump pulse energy) is redistributed
among the total number of holes. The photo-created particles are
approximately $2 \%$ of the background hole density
\cite{Wang_07}. In particular, we consider an excitation by a
monochromatic laser pulse tuned at the energy $E_l$ and having a
pump fluence $P_f$. To estimate the energy $E_{ex}$ transferred
initially from the electromagnetic field to the kinetic energy of
holes and electrons, following \cite{Cywiski_07}, we assume that
the fraction of the laser pulse energy imparted to the holes is
$1/4$ of the photon energy. The total injected kinetic energy is
thus $E_{ex}=n_{ex}^h E_l' \zh$ with
$E_l'=E_l-E_g-(\ze_c^1+\ze_v^1)$ and $\ze_c^1,\ze_v^1$ are the
first eigenvalues of the valence and conduction bands. $n_{ex}^h$
is the density of photo-created particles and $\zh$ is the ratio
of kinetic energy absorbed by the electron gas which can be
estimated within the spherical band approximation as $\zh=
m^{HH}_\parallel/(m_\parallel^c+m_\parallel^{HH})$
\cite{Winkler_book} with $m^{HH}_\parallel$ ($m_\parallel^c$) the
effective mass of the heavy hole (electron) in the parallel
direction of the sample.
\begin{figure}[!t]
\includegraphics[width=.8\textwidth]{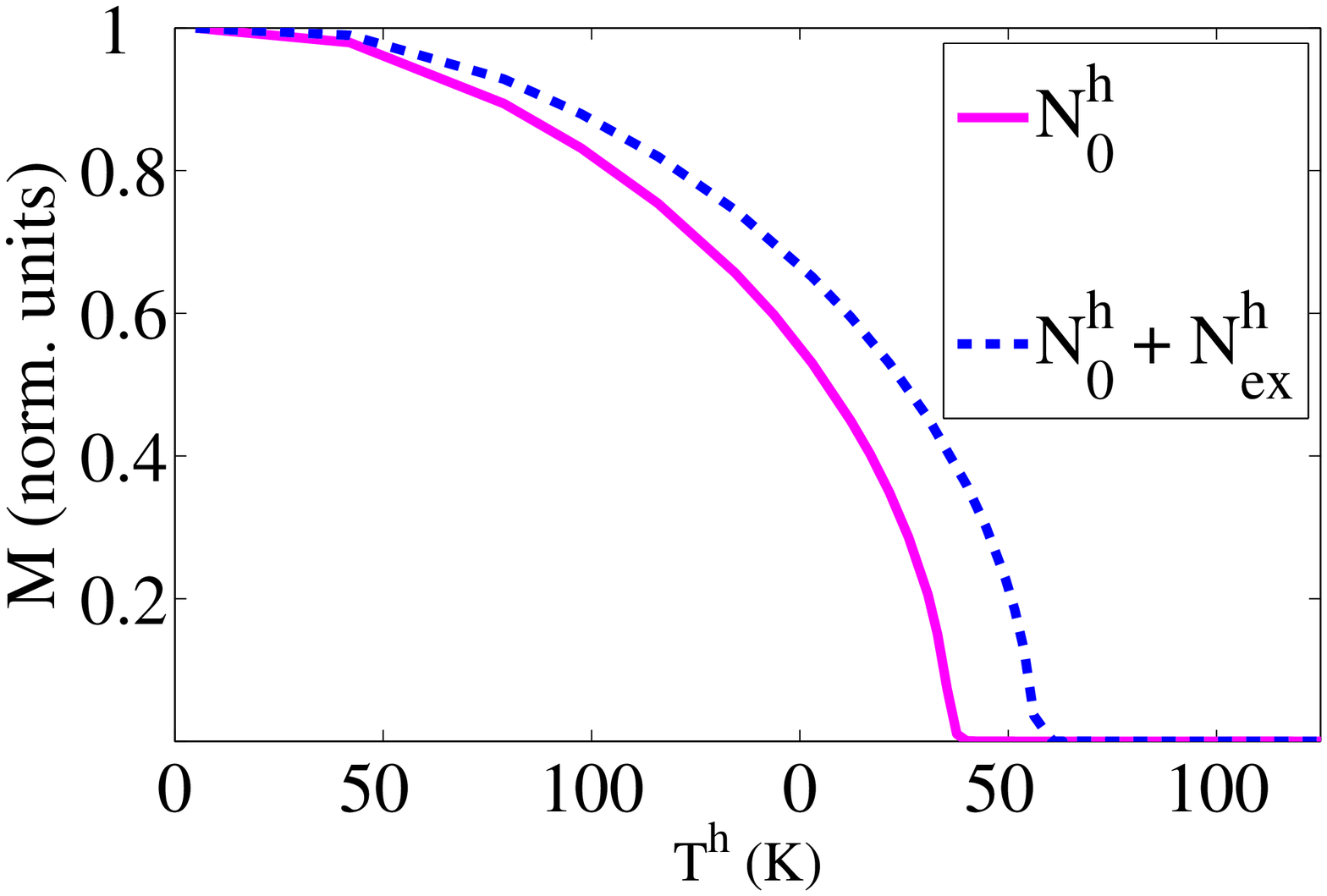}
\caption{Temperature dependence of the normalized equilibrium ion
magnetization for two hole densities. Here $N^h_{ex}= 5\times
10^{19}\textrm{cm}^{-3}.$}\label{fig1}
\end{figure}
In Fig. \ref{fig1} we show the normalized
equilibrium ion magnetization $\mathbf{M}/N^M$ as a function of
the temperature and parameterized by the hole density $N^h$. In
the figure, $N_0^h$ indicates the initial density of holes,
$N_{ex}^h$ the excess of holes excited by the laser pulse, and
$N^h= N_0^h + N_{ex}^h$.
After the laser excitation, the magnetic impurities strongly
interact with the out-of-equilibrium hole gas by means of the
$\mathcal{H}_{pd}$ exchange interaction which redistributes the
spin polarization from one system to another while conserving the
total spin magnetization. Meanwhile, the itinerant hole spin is
efficiently dissipated through spin-orbit interactions
($\tau_{SO}\approx 100$ fs) and relaxation of the total
magnetization can be observed. Short-time spin relaxation of the
holes is therefore an essential ingredient for explaining the
observed time-dependent changes of the magnetization in
ferromagnetic semiconductors \cite{Wang_05, Cywiski_07}.

By means of a standard relaxation model, we include both the
spin-orbit mechanism (or any other mechanisms leading to the
hole-spin relaxation) and the other thermalization effects, such
as the cooling of the kinetic energy of the excited holes driven
by the phonons, and the radiative recombination of the
electron-hole pairs. The corresponding equations read as follow
\begin{eqnarray}
\left.\dpp{n^h_s}{t}\right|_{so}&=&\frac{n^h_s-\overline{n^h_s}}{\tau_{SO}}\label{rela EY}\\
\dpp{N^h}{t}&=&\frac{N^h-N^h_{0}}{\tau_{RR}}\label{rela GR} \\
\dpp{T^h}{t}&=& \frac{T^h-T^L}{\tau_L}\label{rela T}
\end{eqnarray}
where $T^h(t)$ and $T^L$ are the temperatures of the holes and the
lattice, $\overline{n^h_s}(n^M_m,T^h)$ is the self-consistent
quasi-static equilibrium hole spin distribution computed from the
Zener-type model of \cite{Kim_06}, which depends parametrically on
the time-dependent ion magnetization $\mathbf{M}$. The temperature relaxation rate
$\tau_L^{-1}=\tau_{OP}^{-1}+\tau_{AP}^{-1}$ takes into account
both the acoustic phonon scattering with $\tau_{AP}=200$ ps and
the optical phonon scattering with $\tau_{OP} =1 $ ps for $T^h
> 50$ K and $\tau_{OP} =\infty$ for $T^h < 50$ K \cite{Shah_book}. Eq. (\ref{rela GR})
takes into account the radiative recombination process
characterized by a relaxation time $\tau_{RR}=400$ ps
\cite{Shah_book}.
\begin{figure}[!t]
\includegraphics[width=0.6\textwidth]{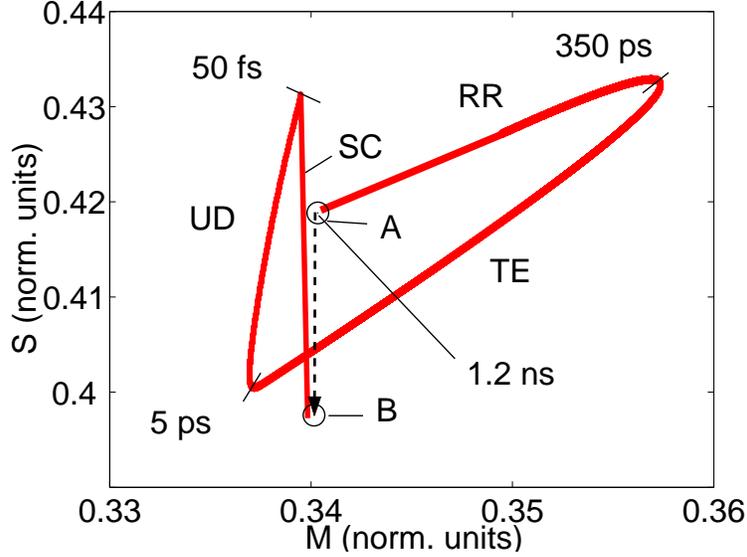}
\caption{Evolution of the normalized holes magnetization
($\overline{S}$) and impurities magnetization ($\overline{M}$), in
the $\overline{S}-\overline{M}$ plane, for $T^L=70$
K.}\label{fig2}
\end{figure}

\begin{figure}[!t]
\begin{center}
\includegraphics[width=8cm]{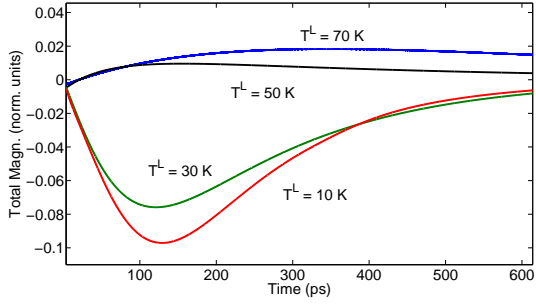}
\caption{Time evolution of the total magnetization for different
lattice temperatures: $T^L=10$ K (red line), $T^L=30$ K (green
line), $T^L=50$ K (black line) and $T^L=70$ K (blue
line).}\label{fig3}
\end{center}
\end{figure}

In Fig. \ref{fig2} we present the time evolution of the normalized
magnetizations, where the vertical axis corresponds to
$\overline{S}=\mathbf{S}/N^h$ and the horizontal axis to
$\overline{M}=\mathbf{M}/N^M$. In agreement with the experiment of
\cite{Wang_07}, we consider a regime of small excitation (laser
pump fluence of 1 $\mu$Jcm$^{-2}$) and a lattice temperature of
$70$ K. The point A represents the initial spin polarization which
is suddenly shifted (instantaneously in our model) to point B.
This is due to the laser excitation which abruptly enhances the
hole density and consequently changes the normalization of
$\overline{S}$, so that
$\overline{S}(0^-)=\frac{\mathbf{S}(0^-)}{N_{0}^h}$ and
$\overline{S}(0^+)=\frac{\mathbf{S}(0^-)}{N^h}$. Our numerical
simulations reveal the presence of different time evolution
regimes: (i) $0 <t < 50$ fs: during this initial phase the
magnetization evolution is nearly coherent (semi-coherent regime
SC in Fig. \ref{fig2}). Indeed, since the photoexcited holes
experience efficient spin-flip scattering with the localized Mn
magnetic moments, a net spin polarization is transferred from the
ion impurities to the holes leading to a significant increase of
the hole spin polarization. Correspondingly, due to the large
difference in densities between the two populations, only a small
decrease of the ion magnetization is observed; (ii) 50 fs $< t< 5$
ps: the nonequilibrium hole spin polarization is efficiently
dissipated via the spin-orbit coupling, which leads to a net
decrease of the total spin magnetization (see also Fig.
\ref{fig3}). During this ultrafast demagnetization regime (UD) the
kinetic temperature of the excited holes is still high; (iii) 5 ps
$<t< 350$ ps: the hole distribution loses its energy via
carrier-phonon scattering and the hole temperature decreases over
the time scale $\tau_{L}$. When the Curie temperature is reached,
the holes and ions spins begin to align, which allows the system
to recover a ferromagnetic order. Since the total number of holes
relaxes to its initial value $N_0^h$ over a slower time scale
$\tau_{RR} \gg \tau_L$, a ferromagnetic state with an excess of
holes can be reached, thus justifying a transient enhancement of
the total magnetization (TE regime); (iv) 350 ps $< t< 1.2$ ns:
finally the radiative recombination of the electron-hole pairs
brings the system back to its initial configuration (RR regime).
\begin{figure}[!t]
\begin{center}
\includegraphics[width=8cm]{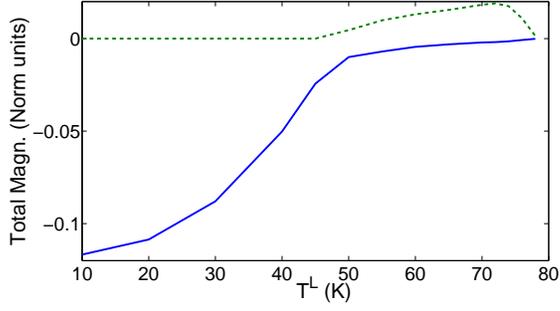}
\caption{Minimum (solid line) and maximum (dashed line) of the
total magnetization for different lattice
temperatures.}\label{fig4}
\end{center}
\end{figure}%

The time evolution of the total magnetization for different
lattice temperatures is depicted in Fig. \ref{fig3}. We see that
the minimum of the total magnetization shifts to shorter times
with increasing lattice temperature, in agreement with
experimental findings. In Fig. \ref{fig4} we plot the excursion of
the total magnetization for different lattice temperature: only
for $45 $ K $<T^L <78$ K an enhancement of the total magnetization
may be observed \cite{Wang_05}.

\begin{figure}[!t]
\begin{center}
\includegraphics[width=8cm]{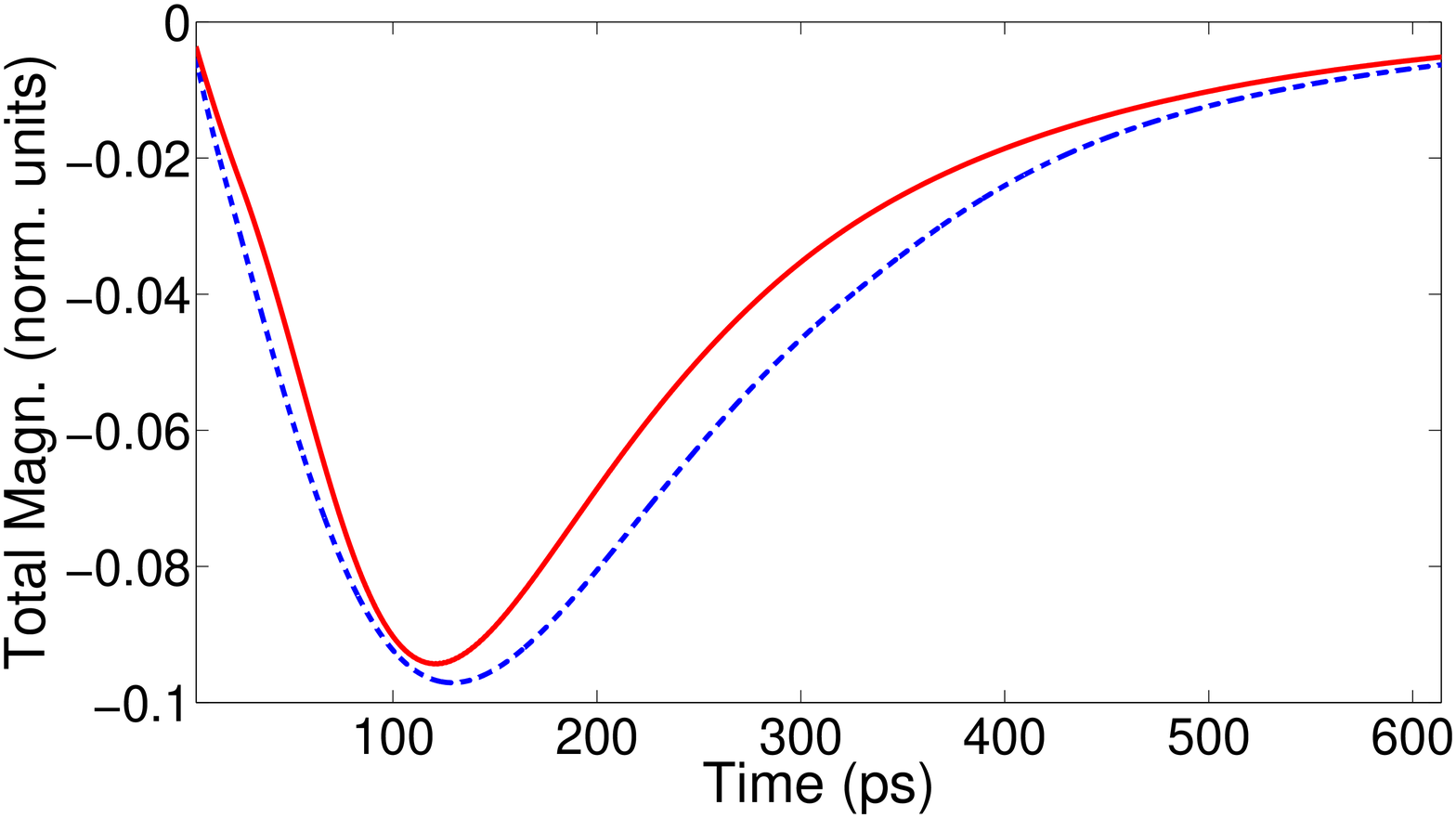}
\caption{Time evolution of the total magnetization at $T^L=10$ K.
Full line: exact formula, Eq. (\ref{Xi}); dashed line:
approximated formula, Eq. (\ref{app_P}).}\label{fig5}
\end{center}
\end{figure}

Finally, in order to validate the approximation of Eq. (\ref{Xi}),
we compare in Fig. \ref{fig5} the time evolution of the total
magnetization obtained by using either the approximate formula
(\ref{app_P}) or by evaluating numerically the integral of Eq.
(\ref{Xi}). As can be clearly seen, a good agreement is obtained,
justifying the use of the simplified expressions (\ref{fin evol
nh}) and (\ref{fin evol nm}).

\section{Conclusion}
In order to describe the strong spin-spin scattering regime
observed in diluted magnetic semiconductors, we have derived a
dynamical model that goes beyond the usual mean-field
approximation. This model is based on the pseudo-fermion formalism
and on a second-order many-particle expansion of the $p-d$
exchange interaction, which is performed in terms of the
single-particle density functions. At this level of description,
this approach is similar to that of Ref. \cite{Cywiski_07}, which
was derived following a different perspective. Numerical
simulations showed that our model is able to reproduce
qualitatively -- and to some extent quantitatively -- the
long-time evolution of the total magnetization after laser
irradiation, as seen in recent experiments \cite{Wang_07}. The
early demagnetization observed in the experiments is explained as
the result of a net flow of polarization from the ions to the
holes, which is subsequently dissipated via spin-orbit coupling.
Thus, the typical demagnetization time scale is mainly determined
by the nonlinear coupling between the ions and holes spins, with a
lower bound given by the spin-orbit time scale, $\tau_{SO} \approx
100~ \rm fs$. The demagnetization process cannot be faster than
$\tau_{SO}$, but can be significantly slower, depending on the
lattice temperature. In addition -- and in contrast to Ref.
\cite{Cywiski_07} -- other slower processes (namely, holes
thermalization and radiative recombination) were also included in
the description, so that the global model encompasses time scales
going from a few tens of femtoseconds to hundreds of picoseconds.

We point out that the methodology developed in this work can be
naturally extended to higher orders by using perturbative
field-theoretic techniques. In particular, we plan to investigate
third order dynamical processes, which were neglected here but may
play an important role in the regime of higher photoexcitation
energy.

Finally, in the model used in this work only the heavy-hole band
contribution to the exchange interaction was taken into account.
The inclusion of a realistic band structure is currently under
study.

\medskip
{\bf \large Acknowledgements}\\
We thank P. Gilliot and J.-Y. Bigot for useful discussions. This
work was partially funded by the Agence Nationale de la Recherche,
contract n. ANR-06-BLAN-0059.


\end{document}